\documentclass[aps,prl,twocolumn,showpacs]{revtex4}

\usepackage{bm}%
\usepackage[colorlinks=true,linkcolor=red]{hyperref}
\usepackage{amsmath}
\usepackage{amsfonts}
\usepackage{amssymb}
\usepackage{verbatim}
\usepackage{epsfig}
\usepackage{graphicx}
\usepackage[sort&compress]{natbib}
\usepackage{times}

\begin{document}

\title{\bf\noindent Large Deviations of the Maximum Eigenvalue 
for Wishart and Gaussian Random Matrices}
\author{Satya N. Majumdar$^1$ and Massimo Vergassola$^2$}
\affiliation{ $^1$ Laboratoire de Physique Th\'{e}orique et Mod\`{e}les
Statistiques (UMR 8626 du CNRS), Universit\'{e} Paris-Sud,
B\^{a}timent 100, 91405 Orsay Cedex (France) \\
$^2$ Institut Pasteur, CNRS URA 2171, F-75724 Paris 15, France.}

\date{\today}

\begin{abstract}
  We present a simple Coulomb gas method to calculate analytically the
  probability of rare events where the maximum eigenvalue of a random
  matrix is much larger than its typical value. The large deviation
  function that characterizes this probability is computed explicitly
  for Wishart and Gaussian ensembles. The method is quite general and
  applies to other related problems, e.g. the joint large deviation
  function for large fluctuations of top eigenvalues. Our results are
  relevant to widely employed data compression techniques, namely the
  principal components analysis. Analytical predictions are verified
  by extensive numerical simulations.

\end{abstract}


\maketitle
Rare events where one of the eigenvalues of a random matrix is much
larger than the others play an important role in data
compression techniques such as the ``Principal Components Analysis''
(PCA).  PCA is a very useful method to detect hidden patterns or
correlations in complex, high-dimensional datasets. A non-exhaustive
list of applications includes image
processing~\cite{Wilks,Fukunaga,Smith}, biological
microarrays~\cite{arrays1,arrays2}, population
genetics~\cite{Cavalli,genetics}, finance~\cite{BP,Burda}, meteorology
and oceanography~\cite{Preisendorfer}. The main idea behind PCA is
very simple.  Consider a rectangular $(M\times N)$ matrix $X$ whose
entries $X_{ij}$ represent some data.  For instance, $X_{ij}$ might
represent examination marks of the $i$-th student ($1\le i\le M$) in
the $j$-th subject (physics, mathematics, chemistry, etc., with $1\le
j \le N$). The product $(N\times N)$ symmetric matrix $W=X^T X$
represents the covariance matrix of the data and it contains
information about correlations. In PCA, one first identifies
eigenvalues and eigenvectors of $W$.  The data are maximally scattered
and correlated along the eigenvector (``principal component'')
associated with the largest eigenvalue $\lambda_{\rm max}$. The
scatter progressively reduces as lower and lower eigenvalues are
considered.  The subsequent step is the reduction of data
dimensionality, achieved by setting to zero those components
corresponding to low eigenvalues.  The {\it rationale} is that
retaining the largest components will preserve the major patterns in
the data and only minor variations are filtered out.

The above description of PCA makes it clear that the efficiency of the
method crucially depends upon the gap between the top eigenvalues and
the ``sea'' of intermediate and small eigenvalues. In particular, the
further is the maximum eigenvalue $\lambda_{\rm max}$ spaced from all
the others, the more effective the projection and the compression
procedure will be. A question then naturally arises: how good is PCA
for random data?  This issue has a twofold interest.  First, in many
situations, the data are high-dimensional and have random
components. Second, random ensembles provide null models needed to
gauge the statistical significance of results obtained for non-random
datasets. To address the question just formulated, one needs to
compute the probability of rare events where the largest eigenvalue
$\lambda_{\rm max}$ has atypically large fluctuations.  The purpose of
this Letter is to provide a simple physical method, based on the
Coloumb gas method in statistical physics, that allows us to compute
analytically the probability of these rare events for a general class
of random matrices.

\begin{figure}
\includegraphics[width=.9\hsize]{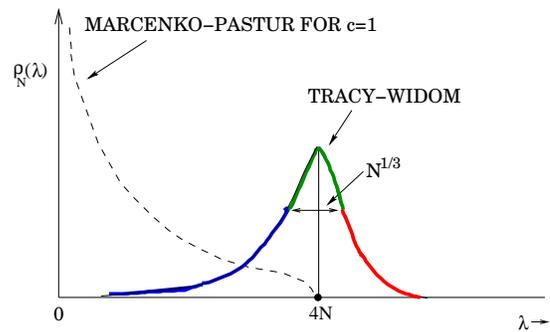}
\caption{The dashed line shows schematically the Marcenko-Pastur
  average density of states for Wishart matrices with the aspect-ratio
  parameter $c\equiv N/M=1$ and the full line is the distribution of the maximum
  eigenvalue $\lambda_{\rm max}$. The PDF is centered around the mean
  value $\langle \lambda_{\rm max} \rangle= 4N$ and {\em typically}
  fluctuates over a scale of width $N^{1/3}$. The probability of
  fluctuations on this scale is described by the known Tracy-Widom
  distribution (green curve).  The red (blue) line on the right (left)
  describes the right (left) large deviation tail of the PDF, which is the
  object of interest in this paper.}
\label{fig:mpc1}
\end{figure}

Let us start by considering Wishart matrices~\cite{Wishart}, which are
directly related to PCA and multivariate statistics ~\cite{Johnstone}.
Wishart matrices are defined via the product $W=X^TX$ of a $(M\times
N)$ random matrix $X$ having its elements drawn independently from a
Gaussian distribution, $P[X]\propto\exp\left[-\frac{\beta}{2}\, {\rm
    tr}(X^\dagger X)\right]$.  The Dyson indices $\beta=1,2$
correspond respectively to real and complex $X$~\cite{Dyson}. Without
any loss of generality, we will assume hereafter that $M\ge N$.  In
addition to the aforementioned PCA applications, Wishart matrices
appear in antenna selection in communication technology~\cite{Mimo},
nuclear physics~\cite{Fyo1}, quantum chromodynamics~\cite{QCD},
statistical physics of directed polymers in random
media~\cite{Johansson} and nonintersecting Brownian
motions~\cite{SMCR}.

The spectral properties of $W=X^TX$ are well-known. For example, $N$
eigenvalues $\{\lambda_i\}$'s of $W$ are nonnegative random variables
with a joint probability density function (PDF)~\cite{James}
\begin{equation}
P\left[\{\lambda_i\}\right] \propto e^{-\frac{\beta}{2}\sum_{i=1}^N\lambda_i}
  \prod_{i=1}^N
  \lambda_i^{\alpha\, \beta/2}
  \prod_{j<k}|\lambda_j-\lambda_k|^\beta\,,
\label{jpdwishart}
\end{equation}
where $\alpha=(1+M-N)-2/\beta$.  This can be written as
$P\left[\{\lambda_i\}\right]\propto \exp\left[-\beta
  E\left(\{\lambda_i\}\right)/2\right]$, with the energy
\begin{equation}
E[\{\lambda_i\}]=\sum_{i=1}^N(\lambda_i-\alpha
\log\lambda_i)-\sum_{j\neq k}\ln |\lambda_j-\lambda_k|\,,
\label{energy}
\end{equation}
coinciding with that of a $2$-d Coulomb gas of charges with
coordinates $\{\lambda_i\}$. Charges are confined to the positive
half-line in the presence of an external linear+logarithmic potential.
The external potential tends to push the charges towards the origin,
whilst the Coulomb repulsion tends to spread them apart. A glance at
(\ref{energy}) indicates that these two competing mechanisms balance
for values of $\lambda$ scaling as $\sim N$. Indeed, from the joint
PDF (\ref{jpdwishart}), one can calculate the average density of
eigenvalues, $\rho_N(\lambda)= \frac{1}{N} \sum_{i=1}^N \langle
\delta(\lambda-\lambda_i)\rangle\approx \frac{1}{N}
f\left(\frac{\lambda}{N}\right)$, with the Marcenko-Pastur
(MP)~\cite{MP} scaling function\,:
\begin{equation}
f(x)= \frac{1}{2\pi x}\, \sqrt{(b-x)(x-a)}\,.
\label{Marcenko-Pastur}
\end{equation}
Here, $c=N/M$ (with $c\le 1$) and the upper and lower edges are
$b=\left(c^{-1/2}+1\right)^2$ and $a=\left(c^{-1/2}-1\right)^2$.  For
all $c<1$, the average density vanishes at both edges of the MP sea.
For the special case $c=1$, we have $a=0$, $b=4$ and the average
density $f(x)= \frac{1}{2\pi}\sqrt{(4-x)/x}$ for $0\le x\le 4$,
diverges as $x^{-1/2}$ at the lower edge (see Fig. \ref{fig:mpc1}).

The MP expression shows that the maximum eigenvalue $\lambda_{\rm
  max}$ has the average value $\langle \lambda_{\rm max}\rangle
\approx b N$ for large $N$. {\rm Typical} fluctuations of
$\lambda_{\rm max}$ around its mean are known to be of
$O(N^{1/3})$~\cite{Johnstone,Johansson}. More specifically,
$\lambda_{\rm max} = b\,N + c^{1/6}\, b^{2/3}\, N^{1/3}\, \chi $,
where $\chi$ has an $N$-independent limiting PDF, $g_{\beta}(\chi)$,
the well-known Tracy-Widom (TW) density~\cite{TW1}. The TW
distribution for $\beta=1,2$ has asymmetric tails~\cite{TW1}
\begin{eqnarray}
g_{\beta}(\chi) &\sim & \exp\left[-\frac{\beta}{24}|\chi|^3\right]
\quad\, {\rm as}\quad \chi\to -\infty\,,
\label{leftTW} 
\\
& \sim & \exp\left[-\frac{2\beta}{3}\chi^{3/2}\right]\quad\, 
{\rm as}\quad \chi\to \infty \label{rightTW}.
\end{eqnarray}
In contrast, the probability of {\em atypically large}, e.g. $\sim
O(N)$, fluctuations of $\lambda_{\rm max}$ from its mean $bN$ are not
captured by the TW distribution. Note that these configurations are
precisely those relevant here, as they are ideally suited for the PCA
to work accurately.

How does the PDF $P(\lambda_{\rm max},N)$ look like for $|\lambda_{\rm
  max}-bN|\gg O(N^{1/3})$ where the TW form is no longer valid?  Using
general large deviation principles, Johansson~\cite{Johansson} proved
that for large fluctuations $\sim O(N)$ from its mean, the PDF
$P(\lambda_{\rm max}=t,N)$ has the form (for large $N$)\,:
\begin{eqnarray}
P(t,N)&\sim & \exp\left[-\beta\, N^2\, \Phi_{-}
\left(\frac{bN-t}{N}\right)\right] \quad t\ll bN \,;
\nonumber \\
& \sim & \exp\left[-\beta\, N\, \Phi_{+}\left(\frac{t-bN}{N}\right)\right]
\quad t\gg bN \,;
\label{rldv1}
\end{eqnarray}
where $\Phi_{\pm}(x)$ are the right (left) rate functions for
the large positive (negative) fluctuations of $\lambda_{\rm max}$. 
The challenge is to explicitly compute their functional
forms. The approach developed for Gaussian matrices~\cite{DM}
allows to compute the left function $\Phi_{-}(x)$~\cite{VMB} but
it does not apply to the right tail. The problem of computing the
right function $\Phi_{+}(x)$ is solved hereafter.  This is
followed by the application of the new method to Gaussian matrices
and further generalizations.

\medskip The starting point of our method to compute $\Phi_{+}(x)$ is
the energy expression (\ref{energy}). The Coulomb gas
physics suggests that when the rightmost charge is moved to the right,
$\lambda_{\rm max}- bN\sim O(N)$, the MP sea is {\it a priori} not
subject to forces capable of macroscopic rearrangements.  Following
this physical picture, the right rate function is determined by the
energy cost in pulling the rightmost charge in the external
potential of the Coulomb gas and the interaction of the charge with
the {\em unperturbed} MP sea. This energy cost for $\lambda_{\rm
  max}=t\gg bN$ can be estimated for large $N$ using
Eq. (\ref{energy})
\begin{equation}
\Delta E (t)= t-\alpha\,\ln (t)-2N\int \ln|t-\lambda|\,
\rho_N(\lambda)\,d\lambda\,,   
\label{ec1}
\end{equation}
where $\rho_N(\lambda)$ is the MP average density of charges and the
integral describes the Coulomb interaction of the rightmost charge
with the MP sea.  This energy cost expression is valid up to an
additive constant, which is chosen such that $\Delta E (t=bN)=0$ since
our reference configuration is the one where $\lambda_{\rm max}=bN$.
For large $N$, we scale $t=zN$, use the MP expression
(\ref{Marcenko-Pastur}) and the energy cost finally takes the form 
\begin{equation}
\frac{\Delta E (z)}{N}=z-\frac{1-c}{c}\ln(z)-
2\int_a^b \ln(z-z')\,f(z')\, dz'\,, 
\label{ec2}
\end{equation}
valid for $z\ge b$ and up to an additive constant.  The probability of
such a configuration is $P(z,N)\propto \exp\left[-\beta\Delta E
  (z)/2\right]$.  Making a shift of variable $z=b+x$, it follows that
$P(t,N)$ for large $N$ and for $t-bN\sim O(N)$ agrees with the large
deviation behavior in Eq. (\ref{rldv1}). Progress is that we also have
derived the explicit expression of the right rate function
$\Phi_{+}(x)$
\begin{eqnarray}
\Phi_{+}(x)= 
\frac{x}{2}&-&\frac{1-c}{2c}\ln\left(\frac{x+b}{b}\right)
\nonumber \\
&-&\int_a^b 
\ln\left(\frac{x+b-x'}{b-x'}\right)\,f(x')\,dx'\,,
\label{rldv-wishart}
\end{eqnarray}
where $x>0$ and the additive constant was chosen to have 
$\Phi_{+}(0)=0$.  The integral can be performed exactly and expressed
as a hypergeometric function. For $c=1$ ($a=0$ and $b=4$), we obtain
\begin{equation}
\Phi_{+}(x)= 
\frac{x+2}{2}-\ln(x+4)+\frac{1}{x+4}\,G
\left(\frac{4}{4+x}\right)\,,
\label{rldv_c1}
\end{equation}
where $G(z)=_3F_2\left[\{1,1,3/2\},\{2,3\}, z \right]$ is a
hypergeometric function (with a lengthy but explicit expression in
terms of elementary functions).  For the sake of comparison, we also
report the simpler expression of the left rate function~\cite{VMB}:
$\Phi_{-}(x)= \ln\left(2/\sqrt{4-x}\right)-x/8-x^2/64$ for $x\ge 0$.

The asymptotics of $\Phi_{+}(x)$ can be easily worked out from
Eq. (\ref{rldv-wishart}).  For large $x$, $\Phi_{+}(x)\sim x/2$
independently of $c$, while the function has a nonanalytic behavior for
small $x$\,:
\begin{equation}
\Phi_{+}(x)\approx \frac{\sqrt{b-a}}{3b}\, x^{3/2}\quad {\rm as} \quad x\to 0.
\label{smallx-wishart}
\end{equation}
This shows that, as $\lambda_{\rm max}-bN \ll O(N)$ 
from the right side, the
PDF of $\lambda_{\rm max}=t$ in Eq. (\ref{rldv1}) behaves as $
\exp\left[-\beta N (\sqrt{b-a}/3b)\,(t/N-b)^{3/2}\right]$. Expressing
the exponent in terms of the TW variable $\chi=c^{-1/6}b^{-2/3}
N^{-1/3}(t-bN)$, we recover exactly the right tail behavior of the TW
density in Eq. (\ref{rightTW}).  Thus, the
large deviation function $\Phi_+(x)$ matches, for small arguments $x$,
the behavior of the TW density at large arguments.  This is quite
consistent with the fact that the scales of the fluctuations for TW
and $\Phi_+(x)$ are $O(N^{1/3})$ and $O(N)$, respectively.  In fact,
our method provides, as a bonus, a physical derivation of
the right tail behavior of the TW density (originally derived through
the Painlev\'e differential equation satisfied by the TW
distribution~\cite{TW1}).

\begin{figure}
\includegraphics[width=.9\hsize]{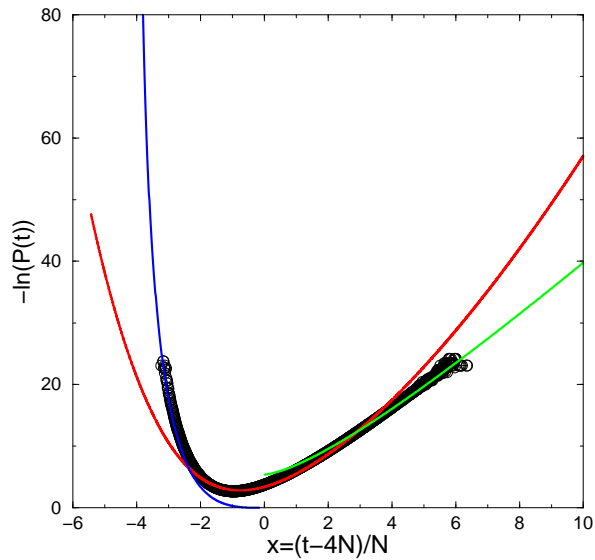}
\caption{Numerical results (circles) for the maximum eigenvalue
  distribution $-\ln P(\lambda_{\rm max}=t,N)$ {\it vs} the scaled
  variable $(t-4N)/N$. Here, $N=10$, Wishart matrices are real
  ($\beta=1$) and $c=1$. Comparisons are with the
  Tracy-Widom distribution (red line) and the exact right (green line)
  and left (blue line) large deviation predictions.}
\label{fig:Wish_c1}
\end{figure}

We confirmed theoretical predictions by extensive numerical
simulations. About $10^{11}$ realizations of real ($\beta=1$) Wishart
matrices of sizes $N=10,\, 26$, $50$, $100$ and with different values
of $c\le 1$ were efficiently generated using the tridiagonal results
in \cite{Edel}. We find very good agreement with our analytical
predictions for the right large deviations. For example, in
Fig. (\ref{fig:Wish_c1}) we present the results for $c=1$ and $N=10$.
MonteCarlo numerical results are compared to the TW density (obtained
by numerically integrating the Painlev\'e equation satisfied by the TW
distribution~\cite{TW1}) and $\Phi_{+}(x)$ in Eq. (\ref{rldv_c1}),
multiplied by $N$. For comparison, we also show the corresponding left
rate function $\Phi_{-}(-x)$ \cite{VMB} multiplied by $N^2$.  It is
clear that, while the numerical data are well described by the TW
density near the peak of the distribution, they deviate considerably
from TW as one moves into the tails, where our large deviation
predictions work perfectly.

\medskip
Our Coulomb gas method is quite general and it can be applied to other
related problems. For example, we can compute the right large
deviation function of $\lambda_{\rm max}$ for Gaussian random matrices.
For the latter, the eigenvalues can be positive or negative with
joint PDF~\cite{Wigner}\,:
\begin{equation}
P\left[\{\lambda_i\}\right] \propto e^{-\frac{\beta}{2}\sum_{i=1}^N\lambda_i^2}
  \prod_{j<k}|\lambda_j-\lambda_k|^\beta\,,
\label{jpdgaussian}
\end{equation}
where the Dyson indices $\beta=1$, $2$ and $4$ correspond to the
orthogonal, unitary and symplectic ensembles. The quadratic nature of
the potential in (\ref{jpdgaussian}), in contrast to the linear term
appearing in (\ref{jpdwishart}), makes that the amplitude of a typical
eigenvalue scales as $\sim \sqrt{N}$. The average density of states
for large $N$ has the scaling form, $\rho_N(\lambda)\approx
\frac{1}{\sqrt{N}}f_{\rm sc}\left(\frac{\lambda}{\sqrt{N}}\right)$,
where the famous Wigner semi-circular law $f_{\rm sc}(x)=
\sqrt{2-x^2}/\pi$ has compact support over $
[-\sqrt{2},\sqrt{2}]$. Thus, $\langle \lambda_{\rm max}\rangle
=\sqrt{2N}$ and typical fluctuations of $\lambda_{\rm max}$ around its
mean are known~\cite{TW1} to be TW distributed over a scale of $\sim
O(N^{-1/6})$. Specifically, $\lambda_{\rm max}= \sqrt{2N} + a_\beta\,
N^{-1/6}\, \chi $, with $a_{1,2}=1/\sqrt{2}$, $a_4= 2^{-7/6}$ and
$\chi$ is a random variable with the TW distribution
$g_{\beta}(\chi)$. Again, the TW form describes the PDF
$P(\lambda_{\rm max}=t,N)$ only in the vicinity of $t=\sqrt{2N}$ over
a small scale of $\sim O(N^{-1/6})$, while deviations from the TW form
appear in the tails. 

Fluctuations of $\lambda_{\rm max}$ over a scale $\sim O(\sqrt{N})$
are described by large deviation functions, analogous to the Wishart
case in Eq. (\ref{rldv1}) but with a different scaling variable
\begin{eqnarray}
P(t,N)&\sim & \exp\left[-\beta\, N^2\, 
\Psi_{-}\left(\frac{\sqrt{2N}-t}{\sqrt{N}}\right)\right]\quad t\ll 
\sqrt{2N}\,;
\nonumber \\
& \sim & \exp\left[-\beta\, N\, \Psi_{+}\left(\frac{t-\sqrt{2N}}{\sqrt{N}}\right)\right]\quad 
t\gg \sqrt{2N}. 
\nonumber
\end{eqnarray} 
\begin{figure}
\includegraphics[width=.9\hsize]{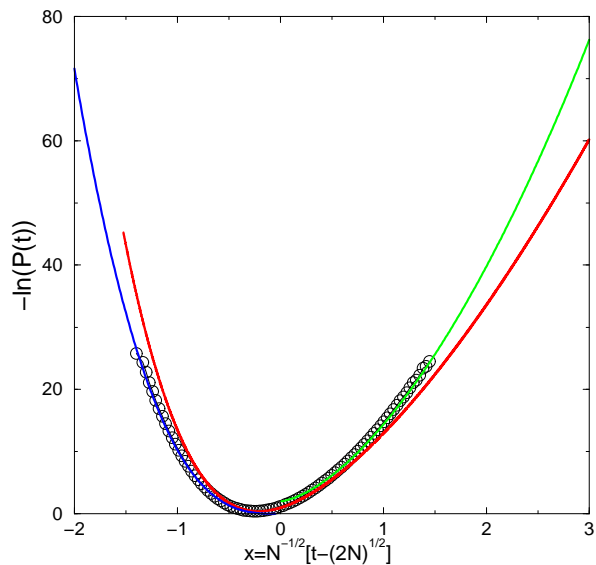}
\caption{Numerical results for the maximum eigenvalue distribution
  (circles) for $N=10$ real ($\beta=1$) Gaussian matrices, compared with
  the Tracy-Widom result (red line) and the exact right (green line)
  and left (blue line) large deviation functions.}
\label{fig:gauss}
\end{figure}
As previously mentioned, the left rate function $\Psi_{-}(x)$ was
recently computed exactly in Ref.~\cite{DM}, but the right rate
function $\Psi_{+}(x)$ was yet unknown. Our Coulomb gas approach
allows to solve this problem as well and gives for $\Psi_{+}(x)$\,:
\begin{equation}
\Psi_{+}(x)= \frac{z^2-1}{2}-\ln(z \sqrt{2})
+\frac{1}{4z^2}\, G\left(\frac{2}{z^2}\right)\,.
\label{rldv2_gaussian}
\end{equation}
Here, $z=\lambda_{\rm max}/N=x+\sqrt{2}$, 
the hypergeometric function $G$ was defined
earlier and the additive constant was chosen to have 
$\Psi_{+}(0)=0$.  The asymptotics of $\Psi_{+}(x)$ can be easily
derived: for large $x$, $\Psi_{+}(x)\sim x^2/2$, while the
non-analytic behavior $\Psi_{+}(x)\approx 2^{7/4}x^{3/2}/3$ holds for
small $x$. Using the TW scaling variable $\chi= \left(\lambda_{\rm
    max}-\sqrt{2N}\right)\, N^{1/6}/ a_\beta$, with $a_{\beta}$
defined after (\ref{jpdgaussian}), it is easy to check that one
recovers the correct TW right tails for all $\beta=1$, $2$ and
$4$. This provides again a physical derivation of the TW right tail.

We have realized simulations for Gaussian matrices with sizes $N=10$,
$25$ and $50$ and for $\beta=1$ and $2$. In Fig. (\ref{fig:gauss}) we
present the data for the PDF of $\lambda_{\rm max}$ (with $N=10$,
$\beta=1$) and compare with the TW form and the exact left function
$\Psi_{-}$ \cite{DM} and right rate function $\Psi_{+}(x)$ derived in
Eq. (\ref{rldv2_gaussian}). As in the Wishart case, the TW form works
well near the peak $t=\sqrt{2N}$, but it fails as we move
into the tails, where the large deviation predictions are quite
accurate.

Our Coulomb gas method lends to further generalizations that we only
briefly mention here. For instance, we can compute the joint
probability distribution for large fluctuations of $n$ top eigenvalues
in Wishart and Gaussian random matrices. If $n\ll N$, the energy will
be given by their mutual charge interactions, the external potentials
and their interaction with the unperturbed MP sea. Integrals are the
same as those computed previously and yield the explicit expression
for the joint PDF. It is also possible to use our method to compute
the large deviation function for fluctuations of the smallest
eigenvalue $\lambda_{\rm min}$ for Wishart matrices with $c<1$.  Note
that the MP sea remains unperturbed (and our method applies) for {\em
  small} fluctuations of $\lambda_{\rm min}$ while the method in
\cite{DM} applies for large fluctuations of $\lambda_{\rm min}$, which
compress the MP sea.

In conclusion, we have presented a new Coulomb gas method to compute
large deviation probabilities of top eigenvalues for a general class
of random matrices. The physical picture that emerges is quite
transparent: when the top eigenvalues are pulled to the right (towards
large values) the Marcenko-Pastur (or Wigner) sea is simply pinched
and the top eigenvalues do not drag all the other eigenvalues. In
other words, no macroscopic rearrangement of the sea occurs and the
top eigenvalues move in the effective potential defined by the
external potential of the Coulomb gas and by the electrostatic potential
generated by the charges in the sea. Our predictions are formally
valid for large $N$ yet our simulations indicate that they work for
moderate $N$ as well. This further adds to the relevance of the large
deviation rate functions derived here to data
compression methods and their applications.

{\bf Acknowledgments} We are grateful to E. Aurell for the invitations to
KTH, where this work was initiated.

\end{document}